\documentclass[12pt]{iopart}
\usepackage[dvips]{graphicx}
\usepackage{multirow}
\usepackage{threeparttable}
\usepackage{setspace}

\usepackage{amsfonts}
\usepackage{mathrsfs}
\usepackage{amsthm}
\usepackage{graphicx}

\begin{document}

%\preprint{}

\title{Research on Solution Space of Bipartite Graph Vertex-Cover by Maximum Matchings}


\author{Wei Wei$^{*}$, Yunjia Zhang, Ting Wang, Baifeng Li, Baolong Niu, Zhiming Zheng}

%\email[]{}
%\thanks{$^{*}$ Corresponding author:weiw@buaa.edu.cn}
\address{LMIB and School of mathematics and systems sciences, Beihang University, 100191, Beijing, China}

\ead{$^{*}$weiw@buaa.edu.cn}


\date{\today}

\begin{abstract}
Some rigorous results and statistics of the solution space of Vertex-Covers on bipartite graphs are
given in this paper. Based on the $K\ddot{o}nig$'s theorem, an exact solution space
expression algorithm is proposed and statistical analysis of the nodes' states is provided.
The statistical results fit well with the algorithmic results until the emergence of
the unfrozen core, which makes the fluctuation of statistical quantities and causes the replica
symmetric breaking in the solutions. Besides, the entropy of bipartite Vertex-Cover solutions
is calculated with the clustering entropy using a cycle simplification
technique for the unfrozen core. Furthermore, as generalization of bipartite graphs,
bipartite core graph is proposed, the solution space of which can also be easily
determined; and based on these results,
how to generate a $K\ddot{o}nig-Egerv\acute{a}ry$ subgraph is studied
by a growth process of adding edges. The investigation of solution space of bipartite
graph Vertex-Cover provides intensive understanding and some insights on the
solution space complexity, and will produce benefit for finding maximal
$K\ddot{o}nig-Egerv\acute{a}ry$ subgraphs, solving general graph Vertex-Cover
and recognizing the intrinsic hardness of NP-complete problems.

\end{abstract}

\maketitle

\section{introduction}

Minimum vertex cover (Vertex-Cover) problem, as one of Karp's 21 NP-complete problems and a classical
graph theoretical computational problem \cite{karp}, has a central status in the research of computational complexity
and attracts the interests of many mathematicians, physicists and computer scientists, which also has a
large number of applications such as immunization strategies in networks \cite{immu}, the prevention
of denial-of-service attacks \cite{denial} and monitoring of internet traffic \cite{moni}. Till now, it is known that
Vertex-Cover problem is NP-complete even in cubic graphs \cite{cubic} and  planar graphs \cite{planar}.
But for bipartite graphs, $K\ddot{o}nig$'s theorem proves a fact that its minimum coverage is
equal to its maximum matchings \cite{konig}.
Besides, for graphs without leaf-removal core, a simple algorithm can help to find a minimal vertex-cover in polynomial time \cite{leaf-removal}.

In statistical mechanics, there are a lot of results obtained on the research of Vertex-Cover problem. In \cite{weight-1,weight-2},
theoretical analysis of Vertex-Cover problem on random graphs is provided to investigate how to solve it efficiently.
In \cite{zhou-1}, a significant structure named long-range frustration, is proposed to explain the strong correlations among the
nodes in Vertex-Cover, and in \cite{zhou-2} and \cite{zhou-3},
the replica symmetric breaking (RSB) analysis on the minimum coverage and its entropy of random graphs
is discussed. Recently, another relationship between the nodes, \emph{mutual-determination}, is investigated to detect the
solution space of Vertex-Cover \cite{wei-1}, which can express the solution space exactly when the graph has no leaf-removal core.
And, a solution number counting algorithm is given based on the solution space expression of Vertex-Cover \cite{wei-2}.
In these research, most results can be proved right only when the graph is of simple structure/topology, such as trees
and graphs with no leaf-removal core, which has to ensure the validity of replica symmetry assumption.
Compared with the Vertex-Cover problem, the satisfiability problem especially $k$-SAT problem
is known as another famous NP-complete problem \cite{sat}, but $k$-XORSAT problem can be solved in polynomial time which undergoes
both the replica symmetry and one step replica symmetric breaking phases (1-RSB) \cite{xorsat}. Can we find a similar easily-solving
subset of Vertex-Cover like that $k$-XORSAT in $k$-SAT?

Based on the $K\ddot{o}nig$'s theorem \cite{konig} and the existing reduced solution graph analysis \cite{wei-2},
the complete solution space of Vertex-Cover on
bipartite graphs will be studied in the paper, and
a polynomial-time algorithm to obtain the exact solution space expression will be provided.
By the results, statistical analysis of the nodes' states in the solution space
is done, which will lose its effectiveness when a structure named unfrozen core emerges.
The existence of the unfrozen core accords with the replica symmetric breaking, as some nodes
can have very large influence on others of the graph, which implicates the long-range correlation
among nodes. Besides, the entropy and the clustering entropy of bipartite graph Vertex-Cover are investigated.
As some generalizations, some easily-solving Vertex-Cover instances are constructed and how to generate
a $K\ddot{o}nig-Egerv\acute{a}ry$ subgraph is discussed.

\section{Vertex-Cover solution space of bipartite graphs}

As is known to all, the classical Vertex-Cover problem is one of the six basic NP-complete problems \cite{planar}.
For arbitrary graphs, till now there is no algorithm in polynomial time to find the minimal vertex-covers,
and to detect its solution space is even harder. In this section,
we will focus on minimal Vertex-Cover on bipartite graphs
which belongs to the $P$ class in computational complexity,
and achieve some rigorous results on the solution space and its statistics.

\subsection{Rigorous solution space expression of Vertex-Cover on bipartite graphs}

In \cite{wei-1}, the reduced solution graph is defined to describe the solution space of Vertex-Cover: double edges are used
to identify the mutual-determinations, in which two ends of the edge
are unfrozen and can determine the states of each other; red or black
nodes are used to identify the positive or negative backbones, which should be always uncovered or covered in the solution space
of Vertex-Cover; single edges retained from the original graph keep the relationship among the nodes by Vertex-Cover.
By the results in \cite{wei-1}, the reduced solution graph can exactly express the whole
solution space of Vertex-Cover when there is no leaf-removal core \cite{leaf-removal} on the given graph. And,
in the following, this result can be generalized to bipartite graphs.

\emph{\textbf{Theorem}}: The solution space of Vertex-Cover for bipartite graphs with $n$ nodes can be exactly
represented by the reduced solution graph in $O(n^2)$ steps, i.e., the states (unfrozen or backbone) of any node can be
easily determined for bipartite graphs.

\textbf{\emph{proof}}: Our inference is based on a very important theorem in graph theory, the $K\ddot{o}nig$'s theorem \cite{konig}:
In any bipartite graph, the number of edges in a maximum matching equals the number of vertices in a minimum vertex-cover.
And, the maximum matching can be solved by an algorithm within polynomial time for general graphs \cite{match},
especially for the bipartite graph it can be solved in $O(n^2)$ steps by the Hungarian algorithm \cite{Hungarian}.

For a bipartite graph $G$, we can first get a maximum matching by the Hungarian algorithm,
e.g.,  $(i_1,j_1)$, $(i_2,j_2)$, $\cdots$, $(i_m,j_m)$.
By the $K\ddot{o}nig$'s theorem, if the maximum matching has $m$ independent edges (edges without common ends),
there should be $m$ covered nodes in the minimal vertex-covers,
which means that each independent edge must have one and only one covered node.
Then, there should be one and only one covered node in each such pair of nodes $(i_k,j_k), k=1,\cdots, m$, which suggests that
the relation in each pair of matched nodes should be mutual-determinations (double edges) in the solution space
of Vertex-Cover; and
no other nodes except $i_1,j_1,i_2,j_2,\cdots,i_m,j_m$ can be covered in the minimal vertex-covers, which suggests that
all the other nodes except those matched have to be always uncovered (positive backbones) in the solution space
of Vertex-Cover.

\begin{figure}
\centering
\includegraphics[width=6.5in]{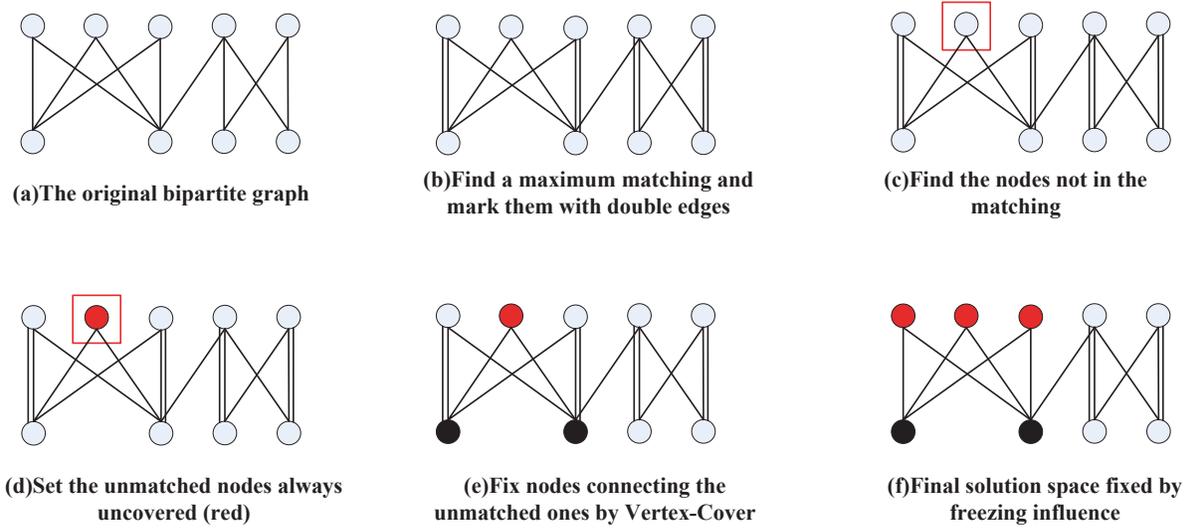}
\caption{An instance for the process of obtaining the reduced solution graph of Vertex-Cover for bipartite graphs.}
\label{fig:bipar-rsg-eg}
\end{figure}

For the obtained graph, if the positive backbones have neighbor nodes on the original graph $G$,
these neighbors should have their states to be always covered (negative backbones); if
any new produced negative backbone, say $i_k$, has one mutual-determination $(i_k,j_k)$ neighbor node $j_k$,
$j_k$ should be always uncovered (positive backbones) by the mutual-determination relation.
This rule is followed by the requirements of Vertex-Cover and should be performed until no
such conditions exist, which costs at most $O(n)$ steps.

\noindent
\begin{tabular}[c]{p{480pt}l}
\textbf{Bipartite Graph Solution Space Expression Algorithm}\\
\hline \hline \textbf{INPUT:}  \texttt{Bipartite graph $G$ } \\
\textbf{OUTPUT:} \texttt{Reduced solution graph $\mathscr{S}(G)$ of $G$}\\
\hline
\textbf{\emph{Freezing-Influence}} \ {(\texttt{Graph $G$,} \texttt{Matching $\mathscr{M}$})}\\
\textbf{begin}\\
\ \ \ \ \ {$\mathscr{S}(G)=$ \texttt{make all matching edges double edges} }\\
\ \ \ \ \ \textbf{while} \ (\texttt{unmatched nodes exist})\\
\ \ \ \ \ \ \ \ \ \ \ {$\mathscr{S}(G)=$ \texttt{make all unmatched nodes uncovered backbones} }\\
\ \ \ \ \ \ \ \ \ \ \ {$\mathscr{S}(G)=$ \texttt{make uncovered backbones' neighbors covered backbone} }\\
\ \ \ \ \ \textbf{while} ({\texttt{any double edge ($i,j$) has node $i$ covered backbone}})\\
\ \ \ \ \ \ \ \ \ \ \ \ \ \ \ \ \ \ \ \ {$\mathscr{S}(G)=$ \texttt{make $j$ uncovered backbone}}\\
\ \ \ \ \ \ \ \ \ \ \ \ \ \ \ \ \ \ \ \ {$\mathscr{S}(G)=$ \texttt{make $j$'s neighbors covered backbone}}\\
\ \ \ \ \ {return($\mathscr{S}(G)$)}\\
\textbf{end}\\
\textbf{\emph{main}}\\
\textbf{begin}\\
\ \ \ \ \ {$\mathscr{M}$=\texttt{maximum matching of} $G$}\\
\ \ \ \ \ {$\mathscr{S}(G)$=\textbf{\emph{Freezing-Influence}}($G$, $\mathscr{M}$)}\\
\textbf{end}\\
\hline
\end{tabular}
\noindent\\

Then, the final obtained graph is the reduced solution graph of $G$. As the whole process is done according to the
$K\ddot{o}nig$'s theorem and the requirements of Vertex-Cover, the reduced solution graph is an exact representation
of the solution space of Vertex-Cover and the total time consumption is at most $O(n^2)$. $\Box$\\

According to the proof of the above theorem, a \emph{Bipartite Graph Solution Space Expression Algorithm} is provided to obtain the
reduced solution graph of a given bipartite graph $G$, and an instance for this process is shown in Fig.\ref{fig:bipar-rsg-eg}.

\subsection{Statistical analysis of bipartite Vertex-Cover nodes' states}

Denote $G=(X_1,X_2,E)$ as a bipartite graph with two independent node sets $X_1$ and $X_2$,
and the edge set $E$ only involves edges connecting $X_1$ and $X_2$.
Define $n_{1}$, $n_{2}$ as the sizes of $X_1$ and $X_2$, $m$ as the size of $E$, $c_{1}=\frac {m} {n_{1}}$, $c_{2}=\frac {m} {n_{2}}$ as the average degrees in $X_1$ and $X_2$, $n=n_1+n_2$ as the total nodes number, and the average degree for the whole graph is $c=2c_1c_2/(c_1+c_2)$. Then, the probability $p$ for the appearance of each edge satisfies $p=\frac {m} {n_{1}\times n_{2}}=\frac {c_{1}} {n_{2}}=\frac {c_{2}} {n_{1}}$, and the degree distributions of $X_1$ and $X_2$ are separately
$p _{1}(k) =C_{n_{2}}^{k}p^{k}(1-p) ^{n_{2}-k}\doteq \frac {c_1 ^{k}} {k!}e^{-c_1}$ and
$p _{2}(k)\doteq\frac {c_2 ^{k}} {k!}e^{-c_2}$ when $n_1, n_2$ are sufficiently large \cite{random}.

Let $Q$ be the size of giant connected component of the bipartite graph, $Q_1$
the node ratio in $X_1$ and $Q_2$ for $X_2$. Thus, one node in $X_1$ belongs
to the giant connected component, if and only if at least one of its neighbors in $X_2$
is in the giant connected component. So, we have
\begin{eqnarray}
% \nonumber to remove numbering (before each equation
  & Q_{1}=1-\sum _{k}p_{1}\left( k\right) \left( 1-Q_{2}\right) ^{k}=1-e^{-c_1\cdot Q_2},  \\
  & Q_{2}=1-\sum _{k}p_{2}\left( k\right) \left( 1-Q_{1}\right) ^{k}=1-e^{-c_2\cdot Q_1}.
\end{eqnarray}
Thus, the size of giant component for the whole graph $G$ is $Q=\frac{Q_1c_2+Q_2c_1}{c_1+c_2}$.

Next, we will calculate the coverage ratio of a random bipartite graph.
Taking the analysis in \cite{coverage}, define $\pi_1$ as the probability of an edge entering
a node in $X_1$ with coverage requirement, i.e., it has not been covered, and $\pi_2$ is the same
but with entering node in $X_2$. Thus, an edge entering a node has coverage requirement only when
the other end of the edge has no coverage requirement from the rest edges, and we have
\begin{eqnarray}
 \pi_1=\sum _{k_1=0}^{\infty}q_{1}(k_1)(1-\pi_2) ^{k_1}, \\
 \pi_2=\sum _{k_2=0}^{\infty}q_{2}(k_2)(1-\pi_1) ^{k_2},
\end{eqnarray}
where $q_i(k)=(k+1)p_i(k+1)/c_i=p_i(k),\ i=1,2$ are the probabilities of an edge entering
a node with degree $k+1$ in $X_i$. Then, the above equations can be simplified by
\begin{equation}\label{equ:pi}
\pi _{1} =e^{-c_{1}\pi _{2}},\ \ \ \ \pi _{2} =e^{-c_{2}\pi _{1}}.
\end{equation}
Using the analysis above,
a node $i$ in $X_1$ should be uncovered only in the following two cases: all its neighbors have
coverage requirements from their edges; all its neighbors except one (say $j$) have
coverage requirements from their edges, and at this time nodes $i$ and $j$ can either be covered.
The analysis is the same for nodes in $X_2$, so we have the coverage ratios in $X_1$ and $X_2$ as
\begin{eqnarray}
\nonumber 
  & x_1=1-\sum_{k=0}^{\infty}p_{1}(k)(1-\pi_{2})^{k}-\sum_{k=1}^{\infty }p_{1}(k)(1-\pi_{2})^{k-1}\frac {k\pi _{2}} {2}&  \\
   & \ =1-e^{-c_{1}\pi_{2}}-\frac{c_{1}\pi_{2}}{2}e^{-c_{1}\pi_{2}},&  \\
   \nonumber 
   &x_2=1-\sum_{k=0}^{\infty}p_{2}(k)(1-\pi_{1})^{k}-\sum_{k=1}^{\infty }p_{2}(k)(1-\pi_{1})^{k-1}\frac {k\pi _{1}} {2}&  \\
   &\ =1-e^{-c_{2}\pi_{1}}-\frac{c_{2}\pi_{1}}{2}e^{-c_{2}\pi_{1}}.&  
\end{eqnarray}
Thus, the coverage ratios of the whole graph $G$ is $x=\frac{x_1*c_2+x_2*c_1}{c_1+c_2}$.

Using similar analysis in \cite{zhou-1,wei-1}, we derive the ratios of backbones and unfrozen nodes in the following.
Let $q_1^+, q_1^-,q_1^0, q_2^+, q_2^-,q_2^0$ be the ratios of
uncovered backbones, covered backbones and unfrozen nodes in $X_1$ and $X_2$ separately, and evidently
$q_{1}^++q_{1}^-+q_{1}^0=1, q_{2}^++q_{2}^-+q_{2}^0=1$.
Then, one new added node should be an uncovered backbone only when all its neighbors
are unfrozen or covered backbones; one new added node should be an unfrozen node only when
there is exactly one uncovered backbone in its neighbors. Thus, we have
\begin{eqnarray}\label{equ:positive}
\nonumber 
& q_{1}^+=\sum_{k=0}^{\infty}p_{1}(k)(1-q_{2}^+)^{k}=e^{-c_{1}q_{2}^+},&\\
& q_{2}^+=\sum_{k=0}^{\infty}p_{2}(k)(1-q_{1}^+)^{k}=e^{-c_{2}q_{1}^+}, &
\end{eqnarray}
\begin{eqnarray}\label{equ:unfrozen}
\nonumber 
& q_{1}^0=\sum_{k=1}^{\infty}p_{1}(k)kq_{2}^+(1-q_{2}^+)^{k-1}=c_1q_2^+e^{-c_1q_2^+},& \\
& q_{2}^0=\sum_{k=1}^{\infty}p_{2}(k)kq_{1}^+(1-q_{1}^+)^{k-1}=c_2q_1^+e^{-c_1q_1^+}.&
\end{eqnarray}
And, the ratios of the uncovered backbones and unfrozen nodes for the whole graph $G$ are
separately $q^+=\frac{q_1^+*c_2+q_2^+*c_1}{c_1+c_2}$ and $q^0=\frac{q_1^0*c_2+q_2^0*c_1}{c_1+c_2}$.
Then, the coverage ratio of the whole graph $G$ is $x=1-q^+-q^0/2$, and it is easy to see that
the results of coverage by equations (6-7) and equations (8-9) are actually the same.
By the shape of equations (\ref{equ:pi}) and (\ref{equ:positive}),
the probability of an edge entering a node with coverage requirement
is the same as the probability of a node being positive backbone.

\begin{figure}
\centering
\includegraphics[width=7in]{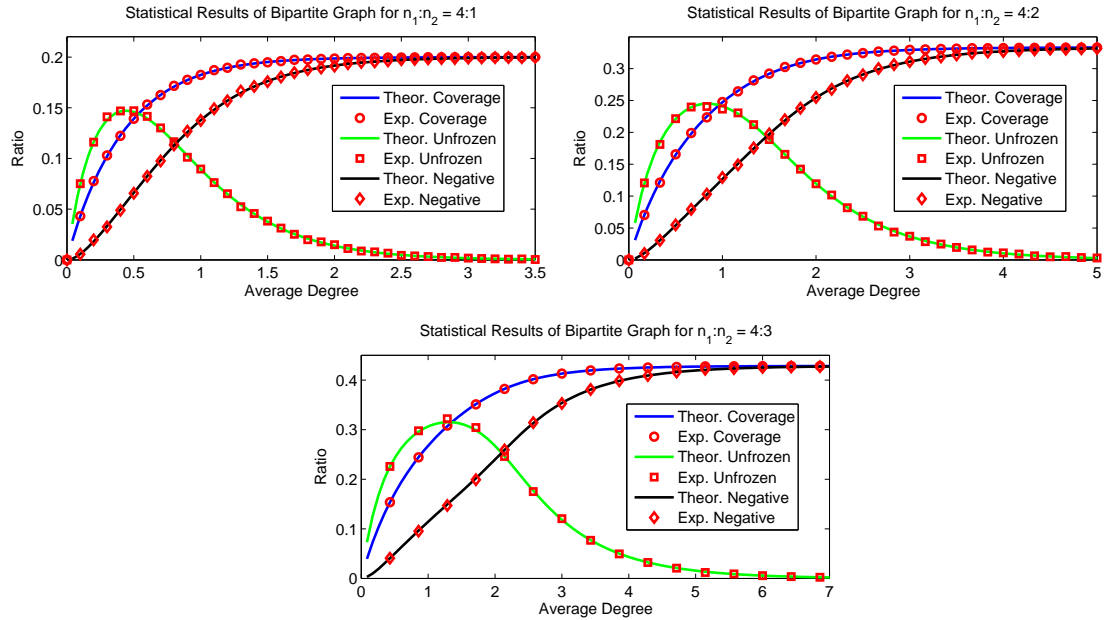}
\caption{The results of minimal coverage, unfrozen nodes and negative backbones for different ratios
of $n_1:n_2=4/1, 4/2, 4/3$ with different average degrees. The theoretical results are shown by
lines with different colors and the experimental results are by data with red symbols. The experimental
results fit very well with those by equations (6-9). All the experimental results are obtained
by 1000 instances with size $n_1+n_2=2000$.}
\label{fig:coverage-unequal}
\end{figure}

For Vertex-Cover of random graphs as we know, when there is a leaf-removal core,
as the existence of long-range correlations \cite{leaf-removal,zhou-1,long-correaltion},
we cannot obtain the exact results of covered and uncovered backbones,
and the key difficulty is how to deal with the correlations among the neighbors of a node.
Fortunately, the correlation among the neighbors is easy to deal with on the bipartite graphs:
if a new added node $i$ has no neighbors of positive backbones,
$i$ being uncovered requires that all its neighbors change to be covered, and it will cause no
confliction among these neighbors.
Assuming two unfrozen neighbors $j_1,j_2$ of $i$ cannot be covered simultaneously,
i.e., $j_1$ being covered causes an influence propagation to affect the state of $j_2$ and requires
$j_2$ being uncovered,
by the organization of the reduced solution graph, the number of nodes in the influence chain between $j_1$ and $j_2$
should be even, e.g., the intermediate nodes $k_1, k_2, \cdots, k_{2n}$,
and we can find that nodes $i, j_1, k_1, k_2, \cdots, k_{2n}, j_2$
form a cycle having $2n+3$ nodes, which conflicts with the organization of
bipartite graph. Thus, we can always have that the unfrozen neighbors of a node can be covered simultaneously for bipartite graphs.
Therefore, the long-range
correlations of Vertex-Cover on bipartite graphs are easygoing and produce no
trouble for determining nodes' states by a node-adding process \cite{zhou-1,wei-1}.
The local evolution of the unfrozen node and negative backbone can be obtained by the same analysis.

In Fig.\ref{fig:coverage-unequal}, the results of minimal coverage, unfrozen nodes and negative backbones
on random bipartite graphs are provided, which fit very well with the analysis of equations in this subsection.
These statistical quantities evolve quite like those on random
graphs \cite{zhou-1}, but the minimal coverage becomes almost unchangeable after some average degrees: for $c_1:c_2=4:1$,
the minimal coverage almost stays to be $x=0.2$ when $c>\sim 3.3$; for $c_1:c_2=4:2$,
the minimal coverage almost stays to be $x=\frac{1}{3}$ when $c>\sim 5.5$; for $c_1:c_2=4:3$,
the minimal coverage almost stays to be $x=\frac{3}{7}$ when $c>\sim8.5$. By the organization of
bipartite graphs, it suggests that after some large average degrees, the minimal coverage, i.e., number of maximum matchings,
should be fulfilled by the nodes in the independent set with smaller size; on the reduced solution graph, after some average degrees,
almost all the nodes in the smaller independent set become negative backbones, and almost all the nodes in the other part
become positive backbones.

%for $c_2=1\cdot c_1$,
%the minimal coverage stays to be $x=\frac{1}{2}$ when $c\gtrsim 9.2$.

\subsection{Fluctuation and unfrozen core of bipartite Vertex-Cover solution space}

\begin{figure}
\centering
\includegraphics[width=6.5in]{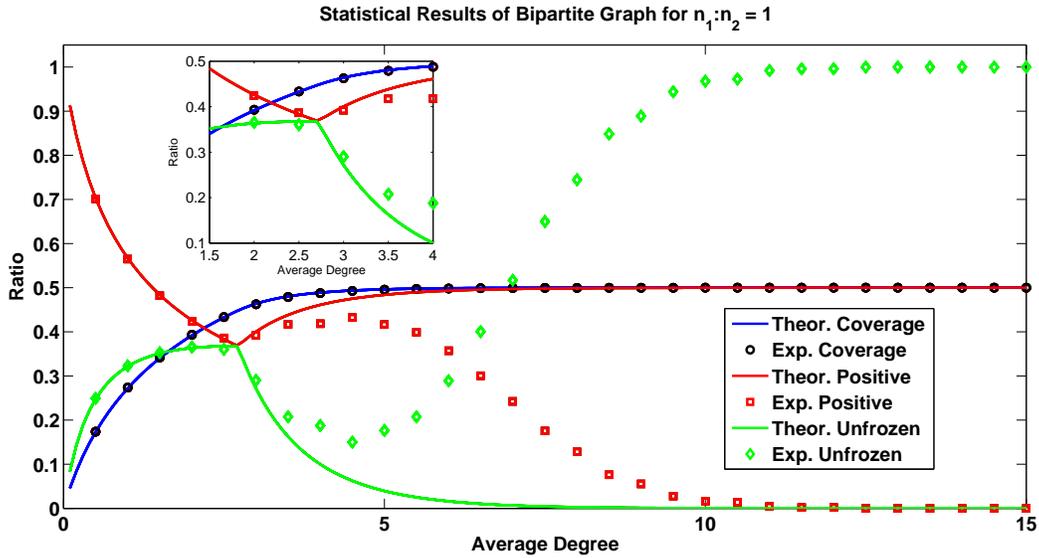}
\caption{The results of minimal coverage, unfrozen nodes and negative backbones for
 $n_1:n_2=1$ with different average degrees, shown separately by blue, green and black lines/symbols.
The experimental results fit well with those by equations (7-11) when $c<e$, but diverge when the average degree
gets larger. All the experimental results are obtained by 1000 instances with size $n_1+n_2=2000$. }
\label{fig:coverage-equal}
\end{figure}

In the above subsection, a statistical analysis for bipartite Vertex-Cover is provided with unequal sizes of two independent sets.
When $c_1:c_2=1:1$, the coverage obtained by these equations also fits very well with the experiments,
but it is surprising that the results of equations (8-9)
deviate from the experimental ones on calculating positive backbones and unfrozen nodes
when the average degree $c>e$,
which are shown in Fig.\ref{fig:coverage-equal}. In this subsection, a new concept, \emph{unfrozen core}, will be introduced
to explain this phenomenon.

Unfrozen core, an organization on the reduced solution graph, is the leaf-removal core for the unfrozen nodes.
In the unfrozen core, some unfrozen nodes couple together and have close connections, which suggests that
there are strong correlations among them. When a new node is added with connection to the unfrozen core,
it being uncovered will lead to a large number of nodes in the unfrozen core to be frozen. Thus,
the unfrozen core is very vulnerable under the local evolution of the node states.

\begin{table}
  \centering
  \begin{tabular}{c||c|c|c|c|c|c|c|c|c|c}
  \hline \hline
  Ave.Deg & $c=0.5$  & $c=1$ & $c=1.5$ &  $c=2$ & $c=2.5$ & $c=3$  &  $c=3.5$   \\
   \hline
  $\rho$(\emph{big ratio})   & 1        & 1     &  1      &      1 & 1       &  0.919 &  0.930   \\
   \hline
  Ave.Deg &  $c=4$ &  $c=4.5$ &  $c=5$  & $c=5.5$  & $c=6$ & $c=6.5$ &  $c=7$  \\
   \hline
  $\rho$(\emph{big ratio})     &  0.914 &  0.877   &  0.860 & 0.796    & 0.702 &  0.636  &  0.458 \\
   \hline
Ave.Deg  & $c=7.5$ & $c=8$  &  $c=8.5$  &  $c=9$ &  $c=9.5$ &  $c=10$ & $c=10.5$ \\
   \hline
  $\rho$(\emph{big ratio})    & 0.380   &  0.231 &  0.164    &  0.095 &  0.068   &  0.030 & 0.021      \\
   \hline
Ave.Deg & $c=11$ & $c=11.5$ &  $c=12$ & $c=12.5$ & $c=13$  &  $c=13.5$  &  $c=14$  \\
   \hline
  $\rho$(\emph{big ratio})   & 0.019  &  0.005   &   0.003 & 0.001    &  0.001  &  0.000    &  0.000      \\
  \hline  \hline
\end{tabular}
  \caption{The sizes $\rho$ of \emph{big ratio} class of positive backbones for different average degrees.
  All the experimental results are obtained by 1000 instances with size $n_1+n_2=2000$.}\label{runtime}
\end{table}

\begin{figure}
\centering
\includegraphics[width=6.5in]{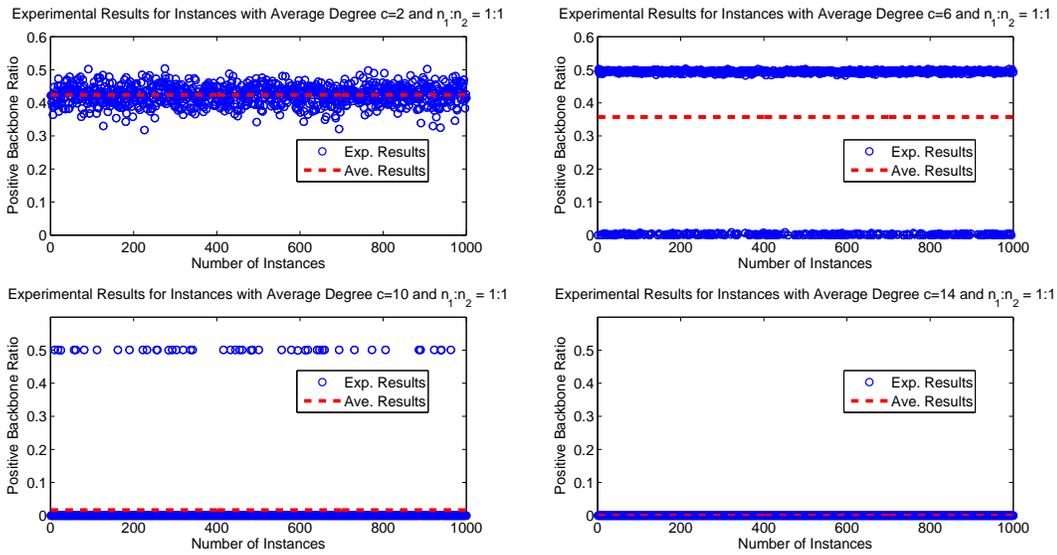}
\caption{The experimental results of positive backbones for different instances
with $n_1:n_2=1$ and average degrees $c=2,6,10,14$. The blue cycles provide
the experimental results of 1000 instances for ratios of positive backbones, and the
red line shows the average over the 1000 instances.
All the instances have size $n_1+n_2=2000$ here. }
\label{fig:fluctuation}
\end{figure}

In Fig.\ref{fig:fluctuation},
an interesting fluctuation phenomenon emerges: when the average degree is small ($c=2$), the positive backbone ratios (the blue cycles)
fluctuate around the average value (the red line) and the average is a typical value;
when the average degree gets larger ($c=6$), the positive backbone ratios split into two different classes,
in which some are around 0.5 (\emph{big ratio}) and the others are around 0 (\emph{small ratio});
when $c=10$, this phenomenon gets clearer and the class of \emph{big ratio} greatly reduces; when $c=14$,
the \emph{big ratio} class disappears. In Table.I, the sizes $\rho$ of the \emph{big ratio} class are provided
to show its evolution.

\begin{figure}
\centering
\includegraphics[width=6.5in]{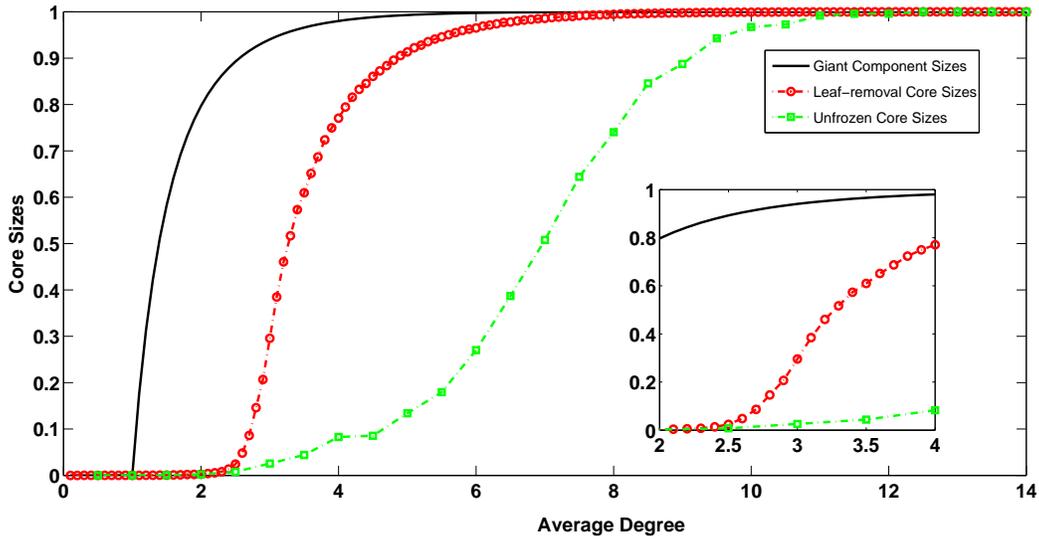}
\caption{The statistical results of giant connected component sizes, leaf-removal core sizes and
unfrozen cores sizes for random bipartite graphs of $n_1:n_2=1$ with different average degrees.
All the experimental results are obtained by 1000 instances with size $n_1+n_2=2000$. }
\label{fig:unfrozen-core}
\end{figure}

In Fig.\ref{fig:unfrozen-core},
the experimental results of sizes for giant component, leaf-removal core and unfrozen core on random bipartite graphs
with $n_1:n_2=1:1$ are provided, in which the emergence of leaf-removal core and unfrozen core is simultaneously.
By the analysis in \cite{wei-1,wei-2}, the emergence of the unfrozen core means the existence of long-range correlations.

For any specific instance, the local evolution of the node states can be exact, which derives from the theorem
of the reduced solution graph for bipartite graphs. Why the statistical results for
positive backbones/unfrozen nodes in equations (8-9) lose their effect?
For statistics in equations (8-9), there is a basic assumption that adding a new node cause little influence
on the macroscopic statistical results. However, when the unfrozen core exists and a new node is added,
if the new node is connected to the unfrozen core, there is a large probability that many nodes' states will be changed
to be frozen in the core, which is shown in Fig.\ref{fig:fluctuation} by the existence of big ratio class of positive backbones.
Indeed, by the results in Fig.\ref{fig:fluctuation} and Fig.\ref{fig:unfrozen-core} when $c>e$,
some instances have unfrozen core, but the others have no unfrozen core as many nodes in the core are
forced to be frozen by some few nodes and the core is broken.
So, one single node in microscopic scale, the influence of which should be neglected in statistics,
may have profound influence on the statistical results in macroscopic scale when the unfrozen core exists. Thus,
the unfrozen core can survive after the node-adding process for some instances, and for other instances,
many of the unfrozen cores get damaged or even disappear, which are determined by the specific
details of the organizations of bipartite graphs. Therefore, the statistical results by equations (8-9) for
the node states are unstable and not exact when the unfrozen core exists after $c>e$.
Similar as the analysis in \cite{xorsat}, the existence of the unfrozen core is quite
like the leaf-removal core in $k$-XORSAT, under which the replica symmetric breaking works.
The long-range correlation among nodes in the unfrozen core can also be
reflected in the core entropy calculated in next section.

\section{Counting the solution number of bipartite Vertex-Cover}

By the above analysis, we can see that there is no \emph{odd-cycle-breaking} \cite{wei-1} in obtaining
the reduced solution graph of bipartite Vertex-Cover, and reduced solution graph for any bipartite graph can be
exactly achieved. To see the organization complexity of the solution space of Vertex-Cover on bipartite graphs,
the entropy $h_s=\frac{log_2 S_n}{n}$ of whole solution space and \emph{core entropy} $h_c=\frac{log_2 S_c}{n}$ \cite{xorsat} are studied in this section,
in which $S_n, S_c$ are separately the total number of minimal vertex-covers and just the number of minimal vertex-covers in the unfrozen core.

For an obtained reduced solution graph (neglecting the backbones), to know the
core entropy, the leaf-removal unfrozen core should be achieved first, which is shown in
Fig.\ref{fig:unfrozen-core} with instances in Fig.\ref{fig:bipar-simplify-eg} (a-c).
However, by the results in Fig.\ref{fig:unfrozen-core}, not all nodes in the
leaf-removal core belong to the unfrozen core, the unfrozen core excludes the backbones and performs
as the intrinsic complexity, and the nodes in the unfrozen core have strong correlations and long-range influence \cite{wei-2}.
By the knowledge of Vertex-Cover on random graph cases \cite{weight-1,weight-2,zhou-1},
solution space (ground states in the terminology of statistical mechanics)
undergoes the replica symmetric breaking phenomenon when the leaf-removal core exists at the average degree $c>e$,
and it should have a replica symmetry phase when there is no leaf-removal core.
So, for the strong correlations among the nodes in the unfrozen core, we can recognize that
the clustering entropy \cite{sat} is caused by the minimal vertex-covers in the unfrozen core, i.e.,
each proper assignment in the unfrozen core leads a cluster of solutions
on the whole reduced solution graph, and the core entropy is indeed the clustering entropy.

To calculate the core entropy $h_c$, we indeed calculate the number of minimal vertex-covers
in the unfrozen core of the reduced solution graph. Here, a technique named \emph{cycle simplification} can be performed:
when there is some cycle with alternatively double and single edges in the unfrozen core,
the nodes on the cycle mutually determine each other, and they can be simplified by
two nodes. In Fig.\ref{fig:bipar-simplify-eg}(a-b), for the six nodes in the unfrozen core,
it is easy to verify that the yellow nodes $i_1, i_2, i_3$ must take the same value in the solution space
and so do the green nodes $j_1, j_2, j_3$, and in Fig.\ref{fig:bipar-simplify-eg}(c)
one yellow node $i$ and one green node $j$ are used to substitute these two classes of nodes, in which
nodes outside the unfrozen core connect $i$ when it originally connects $i_k, k=1,2,3$ or
$j$ when it originally connects $j_k, k=1,2,3$. It is evident that the cycle simplification
does not change the solution number but leads to new reduced solution graph with smaller size.
Then, for the simplified core, the \emph{Vertex-Cover Solution Number Counting
Algorithm} \cite{wei-2} can be used to calculate the clustering entropy $h_c$.
Then, a result for the unfrozen core of the reduced solution graph after
cycle simplification will be given:

\begin{figure}
\centering
\includegraphics[width=6in]{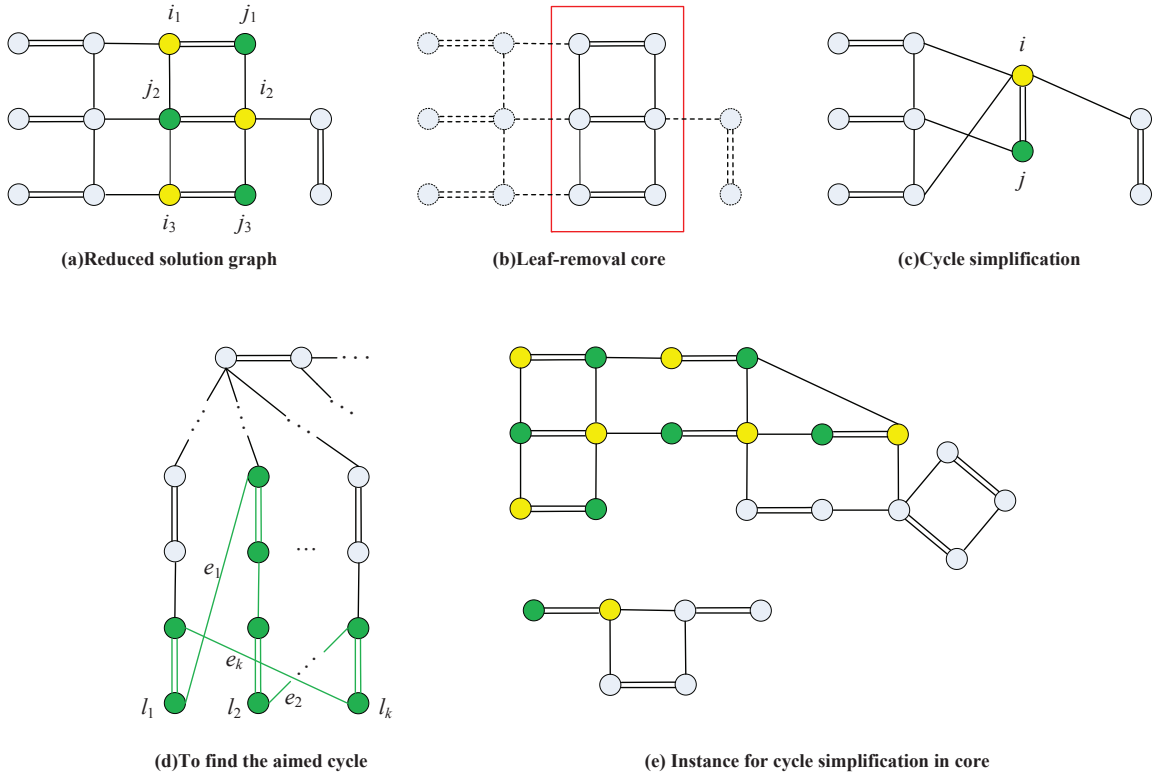}
\caption{Instances for leaf-removal core and cycle simplification of bipartite Vertex-Cover reduced solution graph.
Subgraph (a) is the reduced solution graph, and the green and yellow nodes must take the same values separately.
Subgraph (b) provides the leaf-removal core of (a), and subgraph (c) shows the cycle simplification for (a), in which
the yellow node and green node in (c) correspond to those with the same color in (a).
Subgraph (d) gives a description of how to find an aimed cycle to do cycle simplification,
and the green nodes and edges form the aimed cycle. Subgraph (e) provides
an instance for a core simplification, in which the above one is the original unfrozen core and the below one is the simplified
one. Colored nodes in (e) have the same meaning as those in (a) and (c).}
\label{fig:bipar-simplify-eg}
\end{figure}

\emph{\textbf{Proposition}}: After the cycle simplification operations for $\mathscr{S}(G)$, the new reduced solution graph
$\mathscr{S}'(G)$ has no unfrozen core, i.e., all the unfrozen nodes can be removed after the leaf-removal steps.\\

\emph{\textbf{proof}}: Here, we use proof by contradiction and
assume that there can be leaf-removal unfrozen core $U_c(G)$ after all possible cycle simplification operations.
In the following, we will prove that there must be cycles satisfying the requirement of cycle simplification in $U_c(G)$.
For the unfrozen core $U_c(G)$ of $\mathscr{S}'(G)$, as the connectivity of all the nodes in it,
a spanning tree $T$ \cite{spanning-tree} containing all the double edges of $U_c(G)$ can be easily obtained.
For there is no leaf in $U_c(G)$, any additional single link of $U_c(G)-T$ must produce cycles.

For a pendant $l_1$ of one leaf as the end of one branch of $T$, there must be some edge $e_1$ in $U_c(G)-T$ with end $l_1$, which
connects some other node in $T$ and form a cycle with $l_1$, and then we consider $T\bigcup e_1$.
When the other end (except $l_1$) of edge $e_1$ is one of its ancestors in its branch,
by the bipartite property, the resulted
cycle must be with even number of nodes and satisfy the requirement of cycle simplification, and the contradiction is found.
When the other end of $e_1$ is a node in another branch of $T$,
if the resulted cycle satisfies the requirement of cycle simplification, all the nodes including $l_1$
should be simplified and the contradiction is found; otherwise
consider a new connection $e_2$ of the pendant $l_2$ in the new branch.
For $T\bigcup e_1 \bigcup e_2$, if $l_2$ connects one of its ancestors or some node in the branch of $l_1$ by $e_2$,
then by the bipartite property, the resulted
cycle must satisfy the requirement of cycle simplification and the contradiction is found.
Similar analysis as
$l_1$ can be performed on $l_2$, and if no aimed cycle emerges, we can consider
some $l_3, e_3$ and $T\bigcup e_1 \bigcup e_2\bigcup e_3$, $\cdots$. For this process,
there must be some step that the new pendant $l_k$ connects with some already considered branch (assuming its leaf is $l_i$).
The cycle consisting of $l_i, e_i, l_{i+1}, e_{i+1}, \cdots, l_k, e_k$ should
satisfy the requirement of cycle simplification by the bipartite property, and the contradiction is found.
One schematic view of this process is shown in Fig.\ref{fig:bipar-simplify-eg}(d).
Thus, when there is leaf-removal unfrozen core,
cycle simplification operation can always be performed on it, and
the final resulted graph should have no leaf-removal unfrozen core.
$\Box$

\begin{figure}
\centering
\includegraphics[width=6.5in]{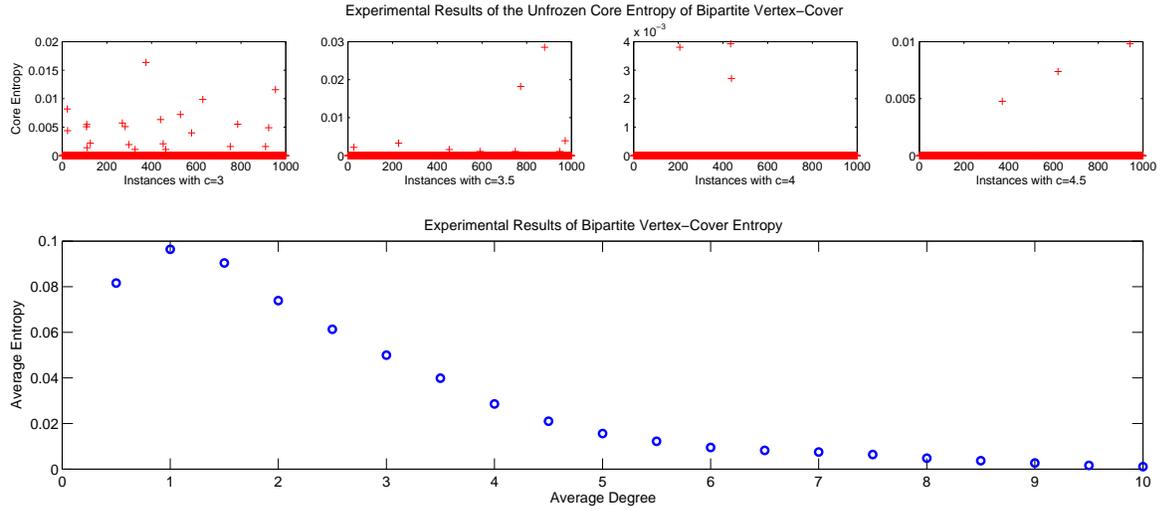}
\caption{The experimental results of entropy and core entropy
of bipartite Vertex-Cover for random bipartite graphs of $n_1:n_2=1$ with different average degrees.
All the experimental results are obtained by 1000 instances with size $n_1+n_2=2000$. }
\label{fig:entropy}
\end{figure}

For the entropy $h_s=\frac{log_2 S_n}{n}$, the number of solutions does not
change after the cycle simplifications,
 the simplified unfrozen core with the removed leaves
should be consider together,
and the calculation of $h_s$ is mainly based
on the \emph{Vertex-Cover Solution Number Counting Algorithm} \cite{wei-2}, which is a complete
algorithm when the reduced solution graph is exact and performs well
when there is no unfrozen core. In Fig.\ref{fig:entropy},
the entropy of random bipartite Vertex-Cover with equal independent sets is provided, which evolves similarly as that
of random graphs \cite{zhou-3}; the core entropy for 1000 instances with average degrees
$c=3,3.5,4,4.5$ are also given and most instances have very low entropy (very few solutions).
As the average degree increases, the unfrozen core increases as that in Fig.\ref{fig:unfrozen-core},
but the solutions in the unfrozen core are very few, which also reveals the
strong correlations among nodes in the unfrozen core.

\section{Easy-solving Vertex-Cover instances and $K\ddot{o}nig-Egerv\acute{a}ry$ subgraph}

\subsection{Vertex-Cover solution space of bipartite core graph}

By the results in \cite{leaf-removal} and \cite{zhou-1}, we know that the minimal vertex-covers
can be easily obtained when there is no leaf-removal core for a graph.
And, by the analysis of the above sections, the solution space of
the bipartite graph can be easily obtained by the maximum matchings.
Thus, if a graph has no leaf-removal core or the leaf-removal core
is a bipartite graph, we name it as \emph{bipartite core graph}, and its solution space expression/reduced
solution graph can be exactly achieved
using the \emph{Mutual-determination and Backbone Evolution Algorithm} \cite{wei-1} and the
\emph{Bipartite Graph Solution Space Expression Algorithm}.\\

Indeed, these two algorithms can be combined simply using the idea of maximum
matching for the \emph{bipartite core graph}: each leaf (a petiole with one pendant
point connecting it) can be viewed as one matching, so the maximum matching number (except the core)
is the same as the number of leaves; for the bipartite leaf-removal core,
the maximum matchings are easily to be obtained. By the minimal coverage,
the covered nodes should be in these two kinds of matchings, and nodes that are not
contained in the matchings must be uncovered backbones.
Thus, for a \emph{bipartite core graph} $G$, we can first find its maximum matchings,
mark them as double edges and unmatched nodes uncovered backbones; then
consider the freezing influence propagation caused by the uncovered backbones, and fix
the states of some nodes like that in the \emph{Bipartite Graph Solution Space Expression Algorithm};
at last, check whether the \emph{odd cycle breaking} operation \cite{wei-1} should be performed,
and do the \emph{freezing influence operation} \cite{wei-1} for the new fixed backbones.
Finally, the obtained compatible graph is the reduced solution graph $\mathscr{S}(G)$, and the algorithm
is given as follows.\\

As the leaves of a \emph{bipartite core graph} may not be bipartite,
e.g., odd cycles can exist (that is why odd-cycle-breaking operations are needed), the \emph{bipartite core graph}
involves not only the bipartite graph and it belongs to the $K\ddot{o}nig-Egerv\acute{a}ry$ graph.
On the contrary, a $K\ddot{o}nig-Egerv\acute{a}ry$ graph cannot always be a \emph{bipartite core graph},
and the structures or characteristics of the $K\ddot{o}nig-Egerv\acute{a}ry$ graph are not very clear till now \cite{np-konig}.
However, if it is known that $G$ is a $K\ddot{o}nig-Egerv\acute{a}ry$ graph, its solution space can be determined
similarly as the \emph{Bipartite Core Graph Solution Space Expression Algorithm}, as all the covered nodes
must be involved in the maximum matchings.\\

\noindent
\begin{tabular}[c]{p{480pt}l}
\textbf{Bipartite Core Graph Solution Space Expression Algorithm}\\
\hline \hline \textbf{INPUT:}  \texttt{Bipartite core graph $G$ } \\
\textbf{OUTPUT:} \texttt{Reduced solution graph $\mathscr{S}(G)$ of $G$}\\
\hline
\textbf{begin}\\
\ \ \ \ \ {\texttt{Find all the leaves by the leaf-removal algorithm} }\\
\ \ \ \ \ {$\mathscr{M}_1$=\texttt{assign one matching in each leaf}}\\
\ \ \ \ \ {$\mathscr{M}_2$=\texttt{maximum matching of bipartite leaf-removal core of} $G$}\\
\ \ \ \ \ {$\mathscr{M}=\mathscr{M}_1\bigcup\mathscr{M}_2$}\\
\ \ \ \ \ {$\mathscr{S}(G)$=\textbf{\emph{Freezing-Influence}}($G$, $\mathscr{M}$)}\\
\ \ \ \ \ {\textbf{while} (Odd-Cycle-Breaking needed)}\\
\ \ \ \ \ \ \ \ \ \ \ \ \ {$\mathscr{S}(G)$=\textbf{\emph{Odd-Cycle-Breaking}}($\mathscr{S}(G)$)}\\
\ \ \ \ \ \ \ \ \ \ \ \ \ {$\mathscr{S}(G)$=\textbf{\emph{Freezing-Influence}}($\mathscr{S}(G)$)}\\
\textbf{end}\\
\hline
\end{tabular}
\noindent\\

\subsection{Generate $K\ddot{o}nig-Egerv\acute{a}ry$ subgraph by bipartite Vertex-Cover solution space}

By the results in \cite{konig,np-konig}, $K\ddot{o}nig-Egerv\acute{a}ry$ graph has its minimum coverage the same as
its maximum matching number, and finding the maximum $K\ddot{o}nig-Egerv\acute{a}ry$ subgraph of a given graph $G(V,E)$
is an NP-complete problem. $K\ddot{o}nig-Egerv\acute{a}ry$ subgraph provides a witness of the minimum coverage
having the maximum matchings' number as a lower bound. However, to find such a maximal $K\ddot{o}nig-Egerv\acute{a}ry$ subgraph
is not always easy as it is NP-complete. Here, we aim to analyze the $K\ddot{o}nig-Egerv\acute{a}ry$ subgraph
using an edge-adding process, in order to see the organization of $K\ddot{o}nig-Egerv\acute{a}ry$ graph
in the viewpoint of reduced solution space. In the following,
one possible way based on our above analysis is provided to obtain a $K\ddot{o}nig-Egerv\acute{a}ry$ subgraph.

We will start the process by a bipartite subgraph $B_G$, which is easy to be obtained from $G$.
Based on $B_G$ (a $K\ddot{o}nig-Egerv\acute{a}ry$ subgraph), we will show how to enlarge it to
keep the $K\ddot{o}nig-Egerv\acute{a}ry$ property by adding new edges and using
the solution space expression of Vertex-Cover. Certainly, any new edge between the two independent sets $X_1$ and $X_2$ for a bipartite graph will
keep the $K\ddot{o}nig-Egerv\acute{a}ry$ property, as it still leads to a bipartite subgraph.
Thus, the following steps are provided as a \emph{Growth Process for $K\ddot{o}nig-Egerv\acute{a}ry$ Subgraph}.

\textbf{\emph{Step 1}} Start with the graph $G(V,\emptyset)$, find a group of maximum matchings of $G$ and add all the matching edges to $G(V,\emptyset)$.

\textbf{\emph{Step 2}} Divide the vertices into two different sets $X_1$ and
$X_2$ by which the two ends of any matching are in different sets,
add all the edges between different sets, and a bipartite subgraph $G_B$ is
obtained.

\textbf{\emph{Step 3}} Using the method of \emph{Bipartite Graph Solution Space Expression Algorithm},
the solution space $\mathscr{S}(G_B)$ of the obtained bipartite graph $G_B$ can be achieved.

Till now, all the edges between $X_1$ and $X_2$ have been added, and then edges in the same sets will be considered.
By the following steps, it is ensured that each obtained subgraph is a $K\ddot{o}nig-Egerv\acute{a}ry$ one and
denoted by $G_{KE}$. Evidently, before these steps, $G_{KE}=G_B$. When a new edge $\in G-G_{KE}$ is considered to be added to $G_{KE}$,
the following steps work:

\textbf{\emph{Step 4}} If one of its two ends is a covered
backbone in $\mathscr{S}(G_{KE})$, the edge can be added directly without changing the solution space of $G_{KE}$.

\textbf{\emph{Step 5}} If both of its two ends are uncovered
backbones in $\mathscr{S}(G_{KE})$, the edge cannot be added into $G_{KE}$ and should be discarded, as it will cause the energy increase.

\begin{figure}
\centering
\includegraphics[width=6in]{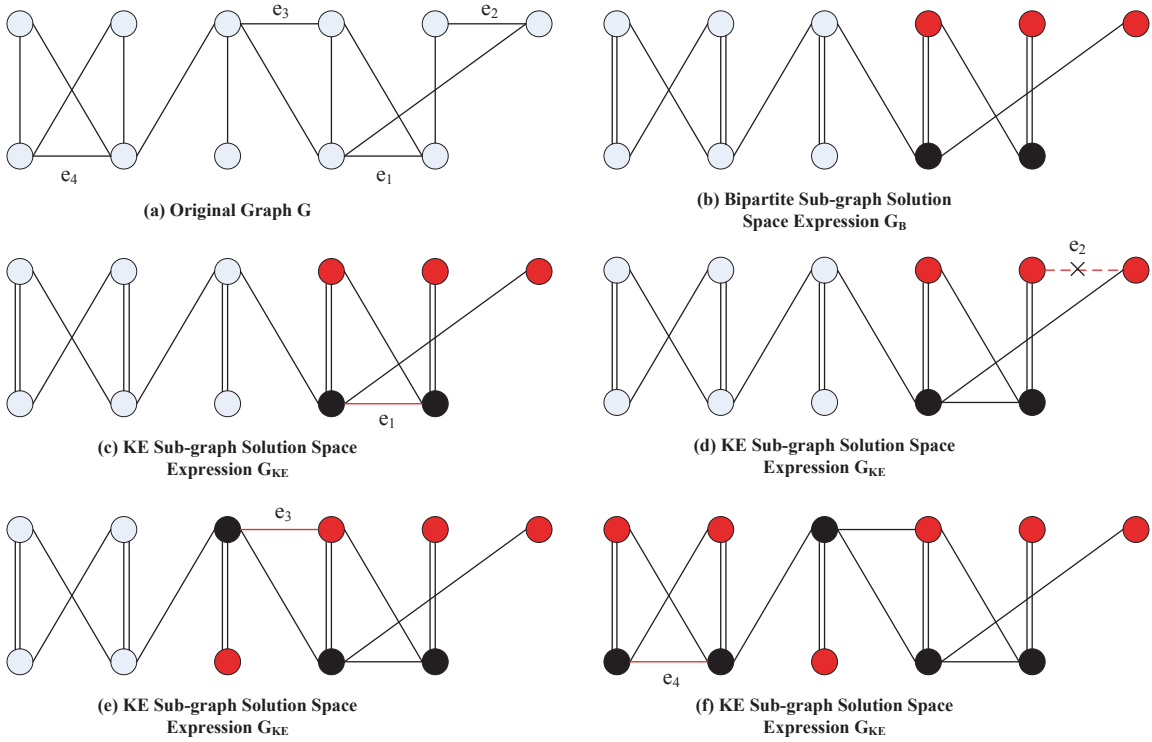}
\caption{The generating process of a $K\ddot{o}nig-Egerv\acute{a}ry$ subgraph
for a given graph $G$. Subgraph (a) is the original graph, and (b) is its bipartite subgraph expressed by the
reduced solution graph. Subgraphs (c-f) reveal the growth process of the $K\ddot{o}nig-Egerv\acute{a}ry$ subgraphs,
which are for the cases in Steps (4-7) correspondingly.}
\label{fig:KE-generating}
\end{figure}

\textbf{\emph{Step 6}} If one of its two ends is an uncovered
backbone with the other unfrozen in $\mathscr{S}(G_{KE})$, by changing the
unfrozen end to covered backbone and doing a \emph{Freezing-Influence} operation on the reduced solution graph of $G_{KE}$,
the edge can be added to $G_{KE}$.

\textbf{\emph{Step 7}} If both of its two ends are unfrozen
in $\mathscr{S}(G_{KE})$, the edge can be added to $G_{KE}$ and an \emph{Odd-Cycle-Breaking} operation should
be perform if necessary.

\textbf{\emph{Step 8}} The above Steps (4-7) are repeated until no new edge in $G-G_{KE}$ can be added,
and the final resulted $G_{KE}$ is obtained.

In Fig.\ref{fig:KE-generating}, an example is provided to show the process of getting a $K\ddot{o}nig-Egerv\acute{a}ry$ subgraph,
which indeed provides a maximal $K\ddot{o}nig-Egerv\acute{a}ry$ subgraph of $G$.
By the steps (1-3) and repeating steps (4-7), the obtained subgraph is a $K\ddot{o}nig-Egerv\acute{a}ry$ one, and its coverage equals to the
maximum matching number. However, the above process cannot always ensure a
maximal $K\ddot{o}nig-Egerv\acute{a}ry$ subgraph of a general graph $G$.
Indeed, inaccuracy arises from the sequence of adding these edges and the \emph{Step 7}.
In \emph{Step 7}, as there may be many choices to do the \emph{Odd-Cycle-Breaking} operation \cite{wei-1}, it may lead to
a contraction of the solution space of $G_{KE}$, and make wrong decisions on discarding more edges in the following steps.
Here, the inaccuracy is a complicated topic for the NP-completeness of maximal
$K\ddot{o}nig-Egerv\acute{a}ry$ subgraph, which will not be studied in this paper.
Certainly, we can record the solution space for different \emph{Odd-Cycle-Breaking} choices by more than one reduced solution graphs,
but when facing many cases of \emph{Step 7}, the space complexity increases exponentially, which leads to the NP-completeness
of determining maximal $K\ddot{o}nig-Egerv\acute{a}ry$ subgraphs. Generally speaking, at least we can make the following conclusion:

\emph{\textbf{Proposition}}: The \emph{Growth Process for $K\ddot{o}nig-Egerv\acute{a}ry$ Subgraph}
keeps the system energy unchanged and the $K\ddot{o}nig-Egerv\acute{a}ry$ property for each step,
and results in a $K\ddot{o}nig-Egerv\acute{a}ry$ subgraph.\\

\section{Conclusion and Discussion}

The solution space description for Vertex-Cover of bipartite graphs
with some further analysis is investigated in this paper.
The exact solution space expression is provided by
the \emph{Bipartite Graph Solution Space Expression Algorithm} for bipartite graphs
and by the \emph{Bipartite Core Graph Solution Space Expression Algorithm} for
bipartite core graphs. The states of the nodes on the bipartite graphs, e.g.,
backbones and unfrozen nodes, are determined by the algorithms, and experiments
are given to support the analysis of their statistics. Specially,
as the emergence of the unfrozen core, the statistical results become not typical,
which implicates the existence of long-range influence of some nodes and the replica symmetric
breaking phenomenon. The entropy and clustering entropy are calculated separately, and some technique
named cycle simplification is proposed to simplify the unfrozen core.
Besides, as extended research, some analysis on the $K\ddot{o}nig-Egerv\acute{a}ry$ (sub)graph
is given to detect the scope of applications for the proposed algorithms.

Though the Vertex-Cover problem on bipartite graph is a \emph{P} problem in computational
complexity, clear understanding of its solution space and nodes' correlations can help us
a lot to recognize the core difficulty of solving Vertex-Cover on general graphs.
Indeed, bipartite graphs can be imbedded as subgraphs to view general graphs and the
modes of nodes' evolution in it can be used for reference to detect the solution organization
mechanism on general graphs. Especially, as the proposed
\emph{Growth Process for $K\ddot{o}nig-Egerv\acute{a}ry$ Subgraph},
the study on bipartite graph can also provide us a schematic view of the
obstacles of achieving a maximal $K\ddot{o}nig-Egerv\acute{a}ry$ subgraph, which is
another NP-complete problem.

\section{Acknowledgment}
This work is supported by the Fundamental Research Funds for the Central
Universities, the National Natural Science Foundation of China (No.11201019),
International Cooperation Project No.2010DFR00700
and Fundamental Research of Civil Aircraft No.MJ-F-2012-04.

\bibliography{basename of .bib file}

\end{document}